\documentstyle[12pt]{article}
\begin{document}
\parskip 10pt plus 1pt
\title{ On the Harmonic Oscillator Realisation of q-Oscillators}
\author{
{\it D.Gangopadhyay, A.P.Isaev}\\
{Laboratory of Theoretical Physics,JINR, Dubna,Russia}\\
}
\date{}
\maketitle
\baselineskip=20pt
\begin{abstract}
The general version of the bosonic harmonic oscillator realisation of 
bosonic q-oscillators is given. It is shown that the currently known 
realisation is a special case of our general solution.

The investigation has been performed at  the Laboratory of Theoretical Physics,
JINR.
\end{abstract}
\newpage

Recently, there has been much interest in quantum Lie algebras which 
first appeared in the investigations of the quantum inverse scattering
problem while studying the Yang-Baxter equations$^{1}$.These quantum
algebras can be considered as a "deformation" of the Lie algebra with 
the numerical deformation parameter $s$ or $q=e^{s}$, such that the 
usual Lie algebra is reproduced in the limit $s\rightarrow 0$ or 
$q\rightarrow 1$. It has been shown that this structure essentially 
connects with quasi-triangular Hopf algebras and its generalisation 
to all simple Lie algebras has been given$^{2}$. There also exists the 
quantum generalisation of the Jordan-Scwinger mapping for $SU(2)_{q}$ 
algebra$^{3}$. Moreover a q-oscillator realisation of many other quantum 
algebras has also been obtained$^{4,5}$. In Ref.4 a harmonic oscillator 
representation of the q-oscillators was also given. The motive of this paper 
is to show that the harmonic oscillator realisation of the q-oscillators 
admits a more general solution than the one currently in vogue$^{4}$.

The basic equations characterising the q-deformed bosonic oscillator system
are
$$aa^{\dagger}- qa^{\dagger}a=q^{-N}~~ ;~~ N^{\dagger}=N\eqno(1)$$
$$[N,a]=-a~;~ Na=a(N-1)\eqno(2)$$
$$[N,a^{\dagger}]=a^{\dagger}~;~ Na^{\dagger}=a^{\dagger}(N+1)\eqno(3)$$
where $a$,  $a^{\dagger}$ are the annihilation and creation operators 
and $N$ is the number operator.

Consider the case when $q$ is complex. Then $(1)$ implies
$$aa^{\dagger}- q^{*}a^{\dagger}a=(q^{*})^{-N}\eqno(4)$$
So from $(1)$ and $(2)$ we get 
$$a^{\dagger}a= {q^{-N}-(q^{*})^{-N}\over q^{*}-q}\eqno(5)$$
Multiplying $(5)$ by $a$ and then commuting $a$ to the right in the right-hand 
side term we obtain

$$aa^{\dagger}= {q^{-N-1}-(q^{*})^{-N-1}\over q^{*}-q}\eqno(6)$$
Substituting $(5)$ and $(6)$ in $(1)$ then gives
$$q^{-N}(q^{*}-q^{-1})= (q^{*})^{-N}(q-(q^{*})^{-1})\eqno(7)$$
Now taking $q=\vert q\vert e^{i\alpha} , q^{*}=\vert q\vert e^{-i\alpha}$
and putting these in $(7)$ we have
$$e^{-i\alpha(N+1)}(\vert q\vert - {1\over \vert q\vert})
= e^{i\alpha(N+1)}(\vert q\vert - {1\over \vert q\vert})\eqno(8)$$
Equation $(8)$ has two solutions
$$\vert q\vert = {1\over \vert q\vert}
~~~i.e.~~~ \vert q\vert=1\eqno(9a)$$
and
$$e^{-2i\alpha(N+1)}=1 
~~~i.e.~~~ \alpha={\pi m\over N+1}\eqno(9b)$$
with $m$ being some integer.The second solution is not appropriate
for us as we consider $q$ as a number and not as an operator. Let us 
take the first solution $(9a)$ {\it viz.} $q=e^{i\alpha}$. Then 
eqs.$(5)$ and $(6)$ can be written as 
$$ a^{\dagger}a=[N]~~~~~,~~~~~ aa^{\dagger}=[N+1]\eqno(10)$$
where $[x]=(q^{x}-q^{-x})/(q-q^{-1})$. It is straightforward to verify that 
$(10)$ is indeed a solution of $(1)$ even if $q$ is real.

Let us address ourselves to determining the representation of the operators
$a$ and $a^{\dagger}$ in terms of ordinary oscillators $\hat a$ and 
$\hat a^{\dagger}$ described by 
$$[\hat a, \hat a^{\dagger}]=1~~,~~\hat N=\hat a^{\dagger}\hat a
=\hat a\hat a^{\dagger}-1$$
$$[\hat N,\hat a]=-\hat a~;~ [\hat N,\hat a^{\dagger}]=a^{\dagger}\eqno(11a)$$
where $\hat N$ is the usual number operator.We now find the solutions for $a,a^{\dagger}$ and $N$ satisfying equations $(1),(2)$ and $(3)$ together with 
$$[\hat N,N]=0~,~[\hat N,a]=-a~,~[\hat N,a^{\dagger}]=a^{\dagger}\eqno(11b)$$
From $(11b)$ one immediately has 
$$N=\Phi(q,\hat N)~,~a=\hat af(q,\hat N)\eqno(12a)$$
with $\Phi$ and $f$ some arbitrary functions at this moment. Reality of $f$
and $(12a)$ then give
$$a^{\dagger}=f(q,\hat N) \hat a^{\dagger}\eqno(12b)$$
Substituting $(12)$ in $(1)$ yields
$$ f^{2}(q,\hat N+1)(\hat N+1)-qf^{2}(q,\hat N)\hat N=q^{-\Phi(q,\hat N)}=q^{-N}\eqno(13)$$
With $q=e^{s}$ this means
$$\Phi(q,\hat N)
=-{1\over s}ln[f^{2}(q,\hat N+1)(\hat N+1)-qf^{2}(q,\hat N)\hat N]\eqno(14)$$
Now from $(2)$ and $(3)$ we have 
$$q^{-N}a=aq^{-N+1}\eqno(15a)$$
$$q^{-N}a^{\dagger}=a^{\dagger}q^{-N-1}\eqno(15b)$$
Putting equation $(13)$ in $(15a)$ results in the functional equation 	
$$F(q,\hat N)({1\over q}+q)-F(q,\hat N -1)-F(q,\hat N +1)=0\eqno(16)$$
where $F(q,\hat N)=f^{2}(q,\hat N)\hat N$.
The same equation is also obtainable from $(15b)$.

In order to solve eq.$(16)$ for $F(q,\hat N)$ note that 
$$F(q,\hat N)\rightarrow\hat N\eqno(17)$$
for $s\rightarrow 0$ or $q\rightarrow 1$. This is simply because 
$f(q,\hat N)\rightarrow 1$  ($a\rightarrow \hat a$) when $q\rightarrow 1$.

Hence we have the following systems of equations:
$$(q+{1\over q})F(q,N)-F(q,N -1)-F(q,N +1)=0\eqno(18a)$$
$$F(1,N)= N~~,~~  \Phi(1,\hat N)=\hat N\eqno(18b)$$
$$ F(q,\hat N+1)-qF(q,\hat N)=q^{-\Phi(q,\hat N)}=q^{-N}\eqno(18c)$$
The last of these equations is essentially equation $(13)$. From $(18c)$ we 
have 
$$F(q,1)=qF(q,0)+q^{-\Phi(q,0)}\eqno(19)$$
From $(18a)$ and $(19)$ we get 
$$F(q,2)=q^{2}F(q,0)+(q+q^{-1})q^{-\Phi(q,0)}$$
A little algebra then leads to the general term 
$$F(q,N)=q^{N}F(q,0)+[N]q^{-\Phi(q,0)}\eqno(20)$$
It is readily verified that $(20)$ satisfies $(18a)$. Hence $(20)$
is the solution of $(18a)$ for arbitrary $F(q,0)$ and $\Phi(q,0)$.
Moreover, note that if $F\equiv\tilde F(q,N)$ is a solution of $(18a)$, then 
$F\equiv\tilde F(q,-N)$ is also a solution.

It is by now obvious that we may write the general solution as 
$$F(q,N)={q^{N}\Phi_{1}(q)-q^{-N}\Phi_{2}(q)\over q-q^{-1}}\eqno(21)$$
where $\Phi_{1,2}$ are arbitrary functions with the restriction that 
$\Phi_{1,2}(1)=1$. Then, using $F(q,\hat N)=f^{2}(q,\hat N)\hat N$ we arrive 
at 
$$f(q,\hat N)=\sqrt{{q^{\hat N}\Phi_{1}-q^{-\hat N}\Phi_{2}\over \hat N(q-q^{-1})}}$$
so that 
$$a=\hat a\sqrt{{q^{\hat N}\Phi_{1}-q^{-\hat N}\Phi_{2}\over \hat N(q-q^{-1})}}$$
$$a^{\dagger}=\sqrt{{q^{\hat N}\Phi_{1}-q^{-\hat N}\Phi_{2}\over \hat N(q-q^{-1})}}\hat a^{\dagger}$$
$$ N=\hat N-{1\over s}ln \Phi_{2}\eqno(22)$$
That the solutions $(22)$ satisfy all the fundamental relations may be easily
established.Choosing $\Phi_{1}=\Phi_{2}=1$ gives the presently known realisation$^{4}$.

Thus we prove that taking into account the additional conditions $(11b)$, the 
representation $(22)$ is the most general.

A similar analysis for fermionic q-oscillators leads to the known result 
$b=\hat b , b^{\dagger}=\hat b^{\dagger}$, and $M=\hat M$ after imposing the 
requirement $M=M^{2}$ for the number operator.

It is our pleasure to thank A.T.Filippov and J.Lukierski for enlightening
discussions.

\end{document}